\documentclass[aip,manuscript=article]{revtex4-1}

\usepackage{graphicx}
\usepackage{dcolumn}
\usepackage{bm}

\usepackage[utf8]{inputenc}
\usepackage[T1]{fontenc}
\usepackage{mathptmx}
\usepackage{etoolbox}

\makeatletter
\def\@email#1#2{%
 \endgroup
 \patchcmd{\titleblock@produce}
  {\frontmatter@RRAPformat}
  {\frontmatter@RRAPformat{\produce@RRAP{*#1\href{mailto:#2}{#2}}}\frontmatter@RRAPformat}
  {}{}
}%
\makeatother

\begin{document}

\title{Optimizing excited states in quantum Monte Carlo: A reassessment of double excitations}

\author{Stuart Shepard}
\email{stushep8989@gmail.com}
\affiliation{MESA+ Institute for Nanotechnology, University of Twente, 7500 AE Enschede, The Netherlands}
\author{Anthony Scemama}
\email{scemama@irsamc.ups-tlse.fr}
\affiliation{Laboratoire de Chimie et Physique Quantiques, Universit\'{e} de Toulouse, CNRS, UPS, 31062 Toulouse, France}
\author{Saverio Moroni}
\email{moroni@democritos.it}
\affiliation{CNR-IOM DEMOCRITOS, Istituto Officina dei Materiali and SISSA Scuola Internazionale Superiore di Studi Avanzati, Via Bonomea 265, I-34136 Trieste, Italy}
\author{Claudia Filippi}
\email{c.filippi@utwente.nl}
\affiliation{MESA+ Institute for Nanotechnology, University of Twente, 7500 AE Enschede, The Netherlands}

\begin{abstract}
Quantum Monte Carlo (QMC) methods have proven to be highly accurate for computing excited states, but the choice of optimization strategies for multiple states remains an active topic of investigation. In this work, we revisit the calculation of double excitation energies in nitroxyl, glyoxal, tetrazine, and cyclopentadienone, exploring different objective functionals and their impact on the accuracy and robustness of QMC. A previous study for these systems employed a penalty functional to enforce orthogonality among the states, but the chosen prefactors did not strictly ensure convergence to the target states. Here, we confirm the reliability of previous results by comparing excitation energies obtained with different functionals and analyzing their consistency. Additionally, we investigate the performance of different functionals when starting from a pre-collapsed excited state, providing insight into their ability to recover the target wave functions.
\end{abstract}

\maketitle

\section{\label{sec:introduction}Introduction}

While mainly employed to study ground-state properties of molecular and condensed systems, quantum Monte Carlo (QMC) methods~\cite{QMC1,QMC2,QMC3}, specifically, variational (VMC) and fixed-node diffusion Monte Carlo (DMC) have recently been shown to be highly accurate approaches also for the computation of excited states of various characters~\cite{Feldt2020,Dash2019,Dash2021,Cuzzocrea2020,Cuzzocrea2022,Shepard2022,Pfau2024,Szabo2024}.
QMC methods provide a stochastic solution to the electronic structure problem, enabling the use of correlated wave functions with explicit dependence on inter-particle coordinates through a Jastrow correlation factor. These methods typically combine Jastrow factors with compact determinantal expansions but can also employ functional forms as complex as actual neural networks to describe the correlated nature of excited states. 

When optimizing the wave function parameters for multiple states of the same symmetry, early calculations employed a state-average approach, where a common set of orbitals and Jastrow parameters were used for all states and optimized to minimize a weighted average of the energies, with the linear coefficients in the expansion ensuring orthogonality among the states~\cite{Filippi2009}. To overcome the potential limitations of these constrained wave functions, state-specific optimization has been used in combination with variance minimization to target individual states~\cite{Otis2023}. However, using variance as the objective function was found to cause problems in some cases, leading to the loss of the target state during minimization~\cite{Cuzzocrea2020}. Alternatively, several approaches to multistate, state-specific optimization have resorted to minimizing objective functionals that include the energies of the relevant states along with a penalty for nonzero overlaps among them~\cite{vmcpen1,vmcpen2,Pathak2021,Shepard2022,Entwistle2023,Wheeler2024}, performing the optimization either one state at a time or simultaneously.

Until recently~\cite{Wheeler2024}, it was unclear how to choose the prefactors in the penalty terms to ensure optimization to the target states. In our previous work~\cite{Shepard2022}, we obtained accurate double excitation energies at the VMC and DMC levels for several challenging molecules by optimizing all states simultaneously using a modification of a penalty functional~\cite{Pathak2021} and relatively small determinantal expansions of balanced quality among the states of interest. However, these calculations were performed with prefactor choices that do not strictly guarantee optimization to the desired states, calling into question the validity of the results and the effectiveness of the approach.

In the present work, we revisit the computation of double excitation energies of nitroxyl, glyoxal, tetrazine, and cyclopentadienone, aiming to reassess the accuracy and robustness of QMC for this class of excitations. With the use of different objective functionals in the optimization of the multiple states, we confirm that the previous claims of accuracy fully hold, reinforcing the reliability of QMC also for excited states of double character.
Furthermore, we explore the ability of different functionals to recover the correct excited states under varying orthogonality prefactor conditions. In particular, we analyze their performance in the extreme case of a pre-collapsed excited state, where the optimization process is especially challenging, gaining better insight into the limitations and strengths of different optimization strategies for excited-state QMC calculations.

\section{\label{sec:methods}Methods}

In the following, we first present the form of the trial wave function. Then, we will focus on three choices of energy functionals with an orthogonality penalty recently employed for the state-specific optimization of excited states in quantum Monte Carlo calculations~\cite{Pathak2021,Shepard2022,Wheeler2024}.

\subsection{Wave Functions}

We employ here QMC trial wave functions of the Jastrow-Slater form,
\begin{equation}\label{WF}
\Psi_{I}=J_I\sum_{i=1}^{N_{\mathrm{CSF}}}c_{Ii}C_i[\left\{ \phi \right\}_{I}],
\end{equation}
where $C_i$ are spin-adapted configuration state functions (CSF), $c_{Ii}$ the expansion coefficients, $J_I$ the Jastrow factors, and $\left\{\phi\right\}_I$ the one-particle orbitals. In all calculations presented here, the ground and excited states have the same symmetry, so the CSFs, $C_i$, share the same orbital occupation patterns for all states. Note that each state, $I$, in Eq.~\ref{WF} has its own optimal Jastrow factor, set of orbitals, and expansion coefficients. When each state has its own optimizable Jastrow and orbital parameters, the approach is referred to as a state-specific optimization, as opposed to a state-average optimization where all states share a common Jastrow factor and set of orbitals which are optimized by minimizing a weighted average of the energies of the individual states~\cite{Cuzzocrea2020,Filippi2009}. In a state-average optimization, only the expansion coefficients are state-dependent and ensure orthogonality among the states.

The initial orbitals in the VMC optimization are obtained from a $N_{\mathrm{state}}$-state-average [SA($N_{\mathrm{state}}$)] CASSCF calculation. The SA-CASSCF orbitals are used as the molecular orbital basis for the CIPSI expansions, which are generated in a balanced manner for the states of interest as described in Ref.~\cite{Shepard2022}. The CIPSI expansion coefficients, $c_{Ii}$, are then used as initial coefficients in the determinantal components of the VMC wave functions.
The subsequent state-specific optimization in VMC will lead to a different set of molecular orbitals for each state.

\subsection{Penalty-Based Objective Functions}

Pathak \textit{et al.}~\cite{Pathak2021} proposed the following objective function for the state-specific optimization of a given state $I$,
\begin{equation}\label{eq:obj:pathak}
 O_{\mathrm{<}}\left[\Psi_I\right]=E_{I}\left[\Psi_I\right] + \sum_{J < I}\lambda_{IJ} |S_{IJ}|^2,
\end{equation}
where the normalized overlap, $S_{IJ}$, is given by,
\begin{equation}\label{eq:overlap}
S_{IJ}=\frac{\langle \Psi_I | \Psi_J \rangle}{\sqrt{\langle \Psi_I | \Psi_I \rangle \langle \Psi_J | \Psi_J \rangle}}.
\end{equation}
The minimization of $O_{\mathrm{<}}$ penalizes non-zero overlaps with states lower in energy than the state under consideration. 
Each overlap penalty is weighted by a state-dependent prefactor, $\lambda_{IJ}>0$. The approach proposed in Ref.~\cite{Pathak2021} involves optimizing states one at a time starting from the ground state. Higher energy states are then optimized using the fixed, pre-optimized, wave functions of lower energy states, or anchoring states. It was also proposed that the prefactors, $\lambda_{IJ}$, should have values on the order of $\Delta E_{IJ} = E_I-E_J$. As discussed below, with these objective functions, it is also possible to optimize all states at the same time, obtaining in principle the same optimal wave functions.

In a more recent work by the same group~\cite{Wheeler2024}, 
a more general optimization approach was introduced which employs a single objective function to simultaneously optimize an ensemble of states,
\begin{equation}\label{eq:obj:wheeler}
 O_e[\{ \Psi_I \} ]=\sum_I w_I E_{I}\left[\Psi_I\right] + \lambda \sum_{I}\sum_{J<I} |S_{IJ}|^2,
\end{equation}
with weights, $w_I>0$, and a state-independent prefactor, $\lambda>0$, applied to the orthogonality penalty. To ensure that the target ensemble of states is the minimum of $O_e$, the weights (which sum to unity) must obey the condition $w_J > w_I$ for $E_J<E_I$ and the minimum value for $\lambda$ must be larger than the maximum of the critical value for each pair of states,
\begin{equation}\label{lc}
        \lambda > \lambda_c = \max_{J<I} \left[ \Delta E_{IJ} \frac{w_Iw_J}{w_J - w_I} \right].
\end{equation}
The gradients of $O_e$ were shown to be equivalent to those of the objective functions (Eq.~\ref{eq:obj:pathak}) in the limit of $w_I/w_J\rightarrow 0$ with $E_I>E_J$. In this limit, the prefactors of the penalty must simply satisfy the condition $\lambda_{IJ} > \Delta E_{IJ}$, as suggested in the first paper~\cite{Pathak2021}. 

Wheeler \textit{et al.}~\cite{Wheeler2024} proposed an iterative approach to determine the weights. We add here how these weights can be calculated exactly for a system with a single unique $\Delta E$ (i.e. a 2-state system, a 3-state system with a doubly degenerate excited state, etc. [see the Appendix]). 

In our previous work~\cite{Shepard2022}, we employed an objective function similar to Eq.~\ref{eq:obj:pathak}, namely, involving the energy of a single state but, as opposed to using anchoring states, we optimized all states at the same time to lower the computational cost and benefit from the use of correlated sampling in the optimization. In addition, we modified Eq.~\ref{eq:obj:pathak} to penalize the states for having non-zero overlap with \emph{all} other states, regardless of energy ordering, which translates to one difference in the objective function,
\begin{equation}\label{eq:obj:shepard}
 O_{\mathrm{\neq}}\left[\Psi_I\right]=E_{I}\left[\Psi_I\right] + \sum_{J \ne I}\lambda_{IJ} |S_{IJ}|^2,
\end{equation}
namely, the $\ne$ in the sum of the overlap penalty,
The prefactors, $\lambda_{IJ}$, were chosen to be symmetric with respect to the interchange of $I$ and $J$. Consequently, to exemplify this in the case of the ground state, while the objective function in Eq.~\ref{eq:obj:pathak} is simply,
\begin{equation}
O_{\mathrm{<}}\left[\Psi_1\right]=E_{1}\left[\Psi_1\right],
\end{equation}
 since there are no states lower in energy, Eq~\ref{eq:obj:shepard} becomes
\begin{equation}
O_{\mathrm{\neq}}\left[\Psi_1\right]=E_{1}\left[\Psi_1\right] + \sum_{J\ne 1}\lambda_{1J} |S_{1J}|^2.
\end{equation}
This modification was motivated by the fact that, in circumstances where states are close in energy, their order may change throughout the optimization while, by introducing `$\ne$' in Eq.~\ref{eq:obj:pathak},  the ordering of the states in energy does not need to be known in advance. 
Using this penalty function, we optimized the variational parameters of Jastrow-Slater wave functions (Eq.~\ref{WF}) built with CIPSI expansions and obtained accurate excitations energies of double character for a set of medium-sized molecules which are challenging for high-level coupled cluster methods.

However, despite our successful application of Eq.~\ref{eq:obj:shepard}, its utilization can lead to potential issues in the optimization~\cite{Wheeler2024}. For a symmetric $\lambda_{IJ}$, minimizing the objective functions $O_{\mathrm{\neq}}\left[\Psi_I\right]$ over the states of interest is in fact equivalent to minimizing an ensemble function, $O_e[\{\Psi_I\}]$, where all weights $w_I$ are equal. 
As a result, the states are constrained to be orthogonal while the sum of the energies of the states is minimized. Since the sum of the energies of a set of eigenstates is equal to that of the states resulting from a rotation of the same eigenstates, the ensemble functional with equal weights has a degenerate minimum. In principle, this can result in an arbitrary rotation of the states yielding an arbitrary excitation energy smaller than that given by the eigenvalues. 
The presence of an issue with using equal weights can also be inferred from Eq.~\ref{lc}: for equal weights, $\lambda_c\to\infty$ so that, for a finite $\lambda$, the objective function will in principle not minimize to the target set of states. 

The use of the objective function $O_{\mathrm{\neq}}$ does of course not pose a problem if $\lambda_{IJ}$ is allowed to be non-symmetric. In this case, the equivalence with $O_e$ becomes apparent by inspecting the gradients with respect to the parameters. If we consider only two states, we can rewrite the gradients of $O_{\mathrm{e}}$ with respect to a parameter, $p_1$, of $\Psi_1$ and a parameter, $p_2$ of $\Psi_2$, as
\begin{eqnarray}\label{eq:grad:wheeler}
 \frac{\partial O_{\mathrm{e}}}{\partial p_1} =w_1\frac{\partial}{\partial p_1}\left( E_{1}\left[\Psi_1\right] + \frac{\lambda}{w_1} |S_{12}|^2\right)  \nonumber\\
 \frac{\partial O_{\mathrm{e}}}{\partial p_2} =w_2\frac{\partial}{\partial p_2}\left( E_{2}\left[\Psi_2\right] + \frac{\lambda}{w_2} |S_{12}|^2\right) ,
\end{eqnarray} 
and compare them with the gradients of $O_{\mathrm{\neq}}$,
\begin{eqnarray}\label{eq:grad:shepard}
 \frac{\partial O_{\mathrm{\neq}} \left[\Psi_1\right]}{\partial p_1} =\frac{\partial}{\partial p_1}\left( E_{1}\left[\Psi_1\right] + \lambda_{12} |S_{12}|^2 \right) , \nonumber\\
 \frac{\partial O_{\mathrm{\neq}}\left[\Psi_2\right]}{\partial p_2} =\frac{\partial}{\partial p_2}\left( E_{2}\left[\Psi_2\right] + \lambda_{21} |S_{21}|^2 \right) .
\end{eqnarray}
Since $S_{12} = S_{21}$, the gradients of $O_{\mathrm{e}}$ and $O_{\mathrm{\neq}}$ are equivalent, up to a multiplicative factor, if $\lambda_{12} = \frac{\lambda}{w_1}$ and $\lambda_{21} = \frac{\lambda}{w_2}$.  Therefore, in practice, $\lambda_{IJ}$ in $O_{\mathrm{\neq}}$ should not be symmetric if optimization to the target states is desired, undermining the original motivation of introducing $O_{\mathrm{\neq}}$, namely, not to have to decide \textit{a priori} the ordering of the states.

In the following, we will revisit the results of Ref.~\cite{Shepard2022} employing either $O_{\mathrm{<}}$ or $O_e$. In both cases, we will optimize all states simultaneously as opposed to using anchoring states in combination with $O_{\mathrm{<}}$~\cite{Pathak2021}. 

\section{\label{sec:computational_details}Computational Details}

Calculations on nitroxyl, glyoxal, tetrazine, and cyclopentadienone are identical to those performed in Shepard {\it et al.}~\cite{Shepard2022} in regards to geometry, basis set, and initial trial wave function. From the referenced work, we focus on the smallest CIPSI expansions in the SA-CASSCF orbital basis which were described in the corresponding SI. To reiterate briefly for the present work, 
all calculations use scalar-relativistic energy-consistent Hartree Fock (HF) pseudopotentials~\cite{Burkatzki2007} with the corresponding aug-cc-pVDZ Gaussian basis sets with diffuse function exponents taken from the corresponding all-electron Dunning's correlation-consistent basis sets~\cite{Kendall1992}. The Jastrow, orbital, and CSF parameters of each state are optimized using the stochastic reconfiguration (SR) method~\cite{SR1,SR2}. All VMC calculations are performed with the software package CHAMP~\cite{CHAMP-EU}, all CIPSI calculations with Quantum Package~\cite{Garniron2019,Scemama2019}, and all HF and SA($N_{\mathrm{state}}$)-CASSCF calculations with GAMESS(US)~\cite{Barca2020}.  For more details on the present calculations, refer to Ref.~\cite{Shepard2022} and, for the CIPSI/QMC approach, refer to Refs.~\cite{Dash2019,Cuzzocrea2020,Dash2021,Cuzzocrea2022}.

\section{\label{sec:results}Results}

\subsection{\label{sec:results:compare}Checking Reliability of Previous Results}\label{sec:results:validate}

We revisit the VMC calculations of Shepard \textit{et al.}~\cite{Shepard2022} on the double excitation energies of four prototypical molecules, leveraging reliable variational optimization approaches and reassessing the results. As already discussed, in our previous work, the objective functional $O_{\mathrm{\neq}}$ was minimized using a symmetric $\lambda_{IJ}$, which can in principle be problematic and lead to an incorrect energy separation among the states. Here, we wish to show that the results from Ref.~\cite{Shepard2022} are accurately representing the actual excitation energies by comparing them with those obtained with an appropriate choice of objective function, $O_{\mathrm{<}}$,  and constraints $\lambda_{IJ}$.

For all molecules, we will repeat the optimizations of the smallest CIPSI expansions built on CASSCF orbitals (see SI of Ref.~\cite{Shepard2022}). Even though the use of natural orbitals or larger expansions on CASSCF orbitals improves the excitation energies in VMC (see main text of Ref.~\cite{Shepard2022}), this choice allows us to perform tests on consistently constructed expansions which are limited in size also for glyoxal and tetrazine.
We compare the previous results to those obtained using $O_{\mathrm{<}}$ as the objective functional in Table~\ref{tab:comparing-to-Shepard2022}. Since this functional imposes no constraints on the ground-state optimization, it unequivocally reflects the optimized ground-state wave function, which, according to the variational principle, is the closest to the eigenstate for the given functional form. This objective functional, in turn, ensures that the minimum-energy state orthogonal to the ground state (i.e., the first excited state) corresponds to the next higher eigenstate, provided that $\lambda_{21}$ satisfies the condition $\lambda_{21} > \left( E_2 - E_1 \right)$. 

The reported energies and excitation energies in Ref.~\cite{Shepard2022} all use values of $\lambda_{21}$  that are safely above this minimum limit. To maintain consistency, we adopt the same values in the present calculations, except of course for $\lambda_{IJ}=0$ when $I<J$, as well as the same optimization procedure (i.e.\ number of steps and SR damping factors). The resulting energies and excitation energies are presented in Table~\ref{tab:comparing-to-Shepard2022} for the smallest CIPSI expansions on on CASSCF orbitals of all four molecules investigated in Shepard \textit{et al.}~\cite{Shepard2022}. 

\begin{table*}[!htb]
\caption{VMC total energies ($E$, Ha) and excitation energies ($\Delta E$, eV) obtained in Ref.~\cite{Shepard2022} ($O_{\mathrm{\neq}}$ with symmetric $\lambda_{IJ}$) and in the present work ($O_{\mathrm{<}}$). In all 2-state optimizations, all non-zero entries of $\lambda_{IJ}$ are 1.0~Ha. The same optimization parameters as in Ref.~\cite{Shepard2022} are used.} 
\begin{tabular}{llllcrc}
\hline
Nitroxyl & \multicolumn{2}{c}{$E\left(1 ^1\mathrm{A'}\right)$} & \multicolumn{2}{c}{$E\left(2 ^1\mathrm{A'}\right)$} & \multicolumn{2}{c}{$\Delta E$} \\
$O_{\mathrm{\neq}}$  & \multicolumn{2}{c}{-26.4891(2)} & \multicolumn{2}{c}{-26.3271(2)} & \multicolumn{2}{c}{4.41(1)} \\
$O_{\mathrm{<}}$ & \multicolumn{2}{c}{-26.4896(2)} & \multicolumn{2}{c}{-26.3264(2)} & \multicolumn{2}{c}{4.44(1)} \\[1ex]
Glyoxal & \multicolumn{2}{c}{$E\left(1 ^1\mathrm{A_g}\right)$} & \multicolumn{2}{c}{$E\left(2 ^1\mathrm{A_g}\right)$} & \multicolumn{2}{c}{$\Delta E$} \\
$O_{\mathrm{\neq}}$ & \multicolumn{2}{c}{-44.5729(2)} & \multicolumn{2}{c}{-44.3633(2)} & \multicolumn{2}{c}{5.70(1)} \\
$O_{\mathrm{<}}$ & \multicolumn{2}{c}{-44.5729(2)} & \multicolumn{2}{c}{-44.3630(2)} & \multicolumn{2}{c}{5.71(1)} \\[1ex]
Tetrazine & \multicolumn{2}{c}{$E\left(1 ^1\mathrm{A_{1g}}\right)$} & \multicolumn{2}{c}{$E\left(2 ^1\mathrm{A_{1g}}\right)$} & \multicolumn{2}{c}{$\Delta E$} \\
$O_{\mathrm{\neq}}$ & \multicolumn{2}{c}{-52.1826(2)} & \multicolumn{2}{c}{-52.0008(2)} & \multicolumn{2}{c}{4.95(1)} \\
$O_{\mathrm{<}}$ & \multicolumn{2}{c}{-52.1830(2)} & \multicolumn{2}{c}{-52.0008(2)} & \multicolumn{2}{c}{4.96(1)} \\[1ex]
Cyclopentadienone & \multicolumn{1}{c}{$E\left(1 ^1\mathrm{A_1}\right)$} & \multicolumn{1}{c}{$E\left(2 ^1\mathrm{A_1}\right)$} & \multicolumn{1}{c}{$E\left(3 ^1\mathrm{A_1}\right)$} & \multicolumn{1}{c}{$\Delta E_{12}$} & \multicolumn{1}{c}{$\Delta E_{13}$} & \multicolumn{1}{c}{$\Delta E_{23}$} \\
$^a$$O_{\mathrm{\neq}}$ & -46.7532(2) & -46.5281(2) & -46.4930(2) & 6.13(1) & 7.08(1) & 0.96(1) \\
$^b$$O_{\mathrm{<}}$  & -46.7534(2) & -46.5278(2) & -46.4937(2) & 6.14(1) & 7.07(1) & 0.93(1) \\
\hline
\multicolumn{7}{l}{$^a$ $\lambda_{12,21}=1.5$~Ha, $\lambda_{13,31}= 2.0$~Ha, $\lambda_{23,32}= 1.0$~Ha.}\\
\multicolumn{7}{l}{$^b$ $\lambda_{21}=1.5$~Ha, $\lambda_{31}= 2.0$~Ha, $\lambda_{32}= 1.0$~Ha, with the rest zero.}\\
\end{tabular}
\label{tab:comparing-to-Shepard2022}
\end{table*}

\begin{figure}[htb]
\begin{center}
\includegraphics[width=0.8\textwidth]{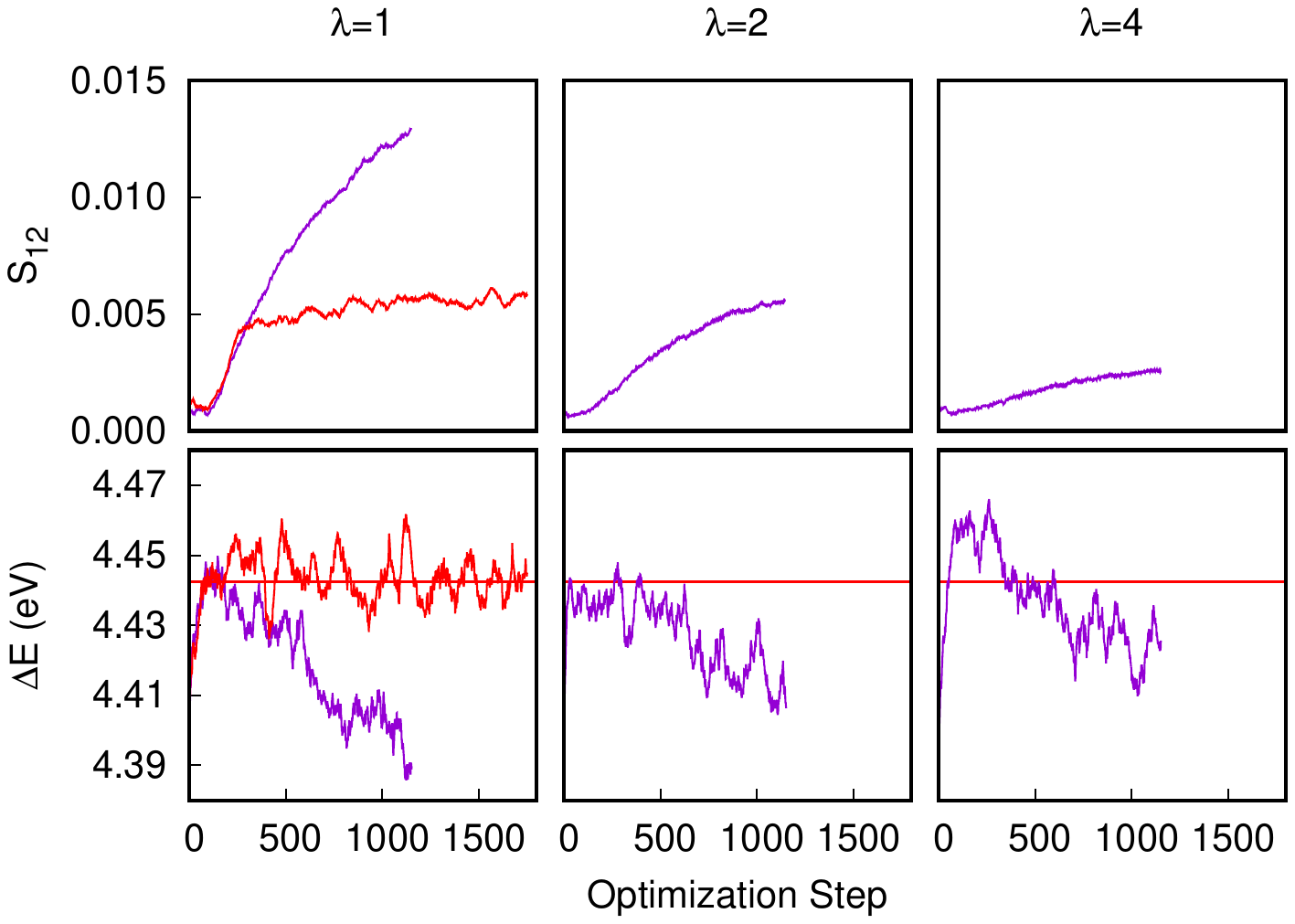}
\caption{\label{fig:validate-overlap} Variational optimization of the ground and excited states of nitroxyl using the objective functions (red) $O_{\mathrm{<}}$ and (purple) $O_{\mathrm{\neq}}$. We plot (top) the overlap between the two states and (bottom) the excitation energy as a function of the optimization step and for increasing values of $\lambda$ factor (a.u.) in the penalty.  }
\end{center}
\end{figure}

Overall, the two sets of results are in good agreement, confirming the reliability of the QMC calculations of Shepard \textit{et al.}~\cite{Shepard2022} on the double excitation energies of these molecules. Notably, the excitation energy in the nitroxyl calculation is in disagreement for the reported precision, the excitation obtained with the $O_{\neq}$ functional being sightly lower (we note that the third state of cyclopentadienone also differs, but with a higher excitation energy, which is not attributable to the issue we are analyzing). Nitroxyl is in fact the only system for which we observe a growing overlap when the $O_{\neq}$ objective function is employed as also shown in Figure~\ref{fig:validate-overlap}. The optimization of the nitroxyl wave functions were however particularly long, allowing us to observe the progressive mixing between the states and the corresponding reduction in the excitation energy, an effect which could have remained hidden in the shorter optimizations for the larger molecules. Increasing the weighting factor, $\lambda$, in the penalty term mitigates the issue but does not entirely eliminate it: as $\lambda$ is increased from 1 to 2 to 4 a.u, state mixing appears at progressively larger step counts. On the other hand, the use of the $O_{\mathrm{<}}$ function in combination with the smallest value of $\lambda$ leads to a plateau in the overlap and prevents the reduction of the excitation energy, confirming the robustness of the approach.

Therefore, in practice, when the $O_{\neq}$ with a symmetric $\lambda_{IJ}$ is employed as in Ref.~\cite{Shepard2022}, the states might remain close to the target ones in optimization runs of typical length if a sufficiently large weight in the penalty is adopted and if the starting trial wave functions are of good quality as in our case. However, in general, this practice will lead to state admixture and an underestimated excitation energy, and one should instead use appropriate constraints in combination with the $O_{<}$ or $O_{\mathrm{e}}$ functionals. That said, the excitation energies obtained with $O_\mathrm{<}$ in Table~\ref{tab:comparing-to-Shepard2022} confirm that the results in Ref.~\cite{Shepard2022} reflect the predictive capability of the QMC/CIPSI approach for the computation of double excitations which was the primary focus of that work.

\subsection{\label{sec:results:testing}Testing Objective Functions with Pre-Collapsed States}

To challenge each approach and further test the prescribed constraints on $\lambda_{IJ}$, we pre-collapse the small CIPSI wave function for the excited state of nitroxyl by optimizing it with no overlap penalty, and then continue the optimization imposing different orthogonality constraints. The lengths and parameters of the SR optimization are identical in all runs, and only the objective functional and penalty prefactor are changed. 
We represent the prefactors as multiples of a critical value, $\lambda = a \lambda_c$, where we assume $\lambda_c$ to take the extrapolated full CI value of 0.159 a.u. (4.32 eV) for the excitation energy of nitroxyl. 

\begin{figure}[htb]
\begin{center}
\includegraphics[scale=0.6]{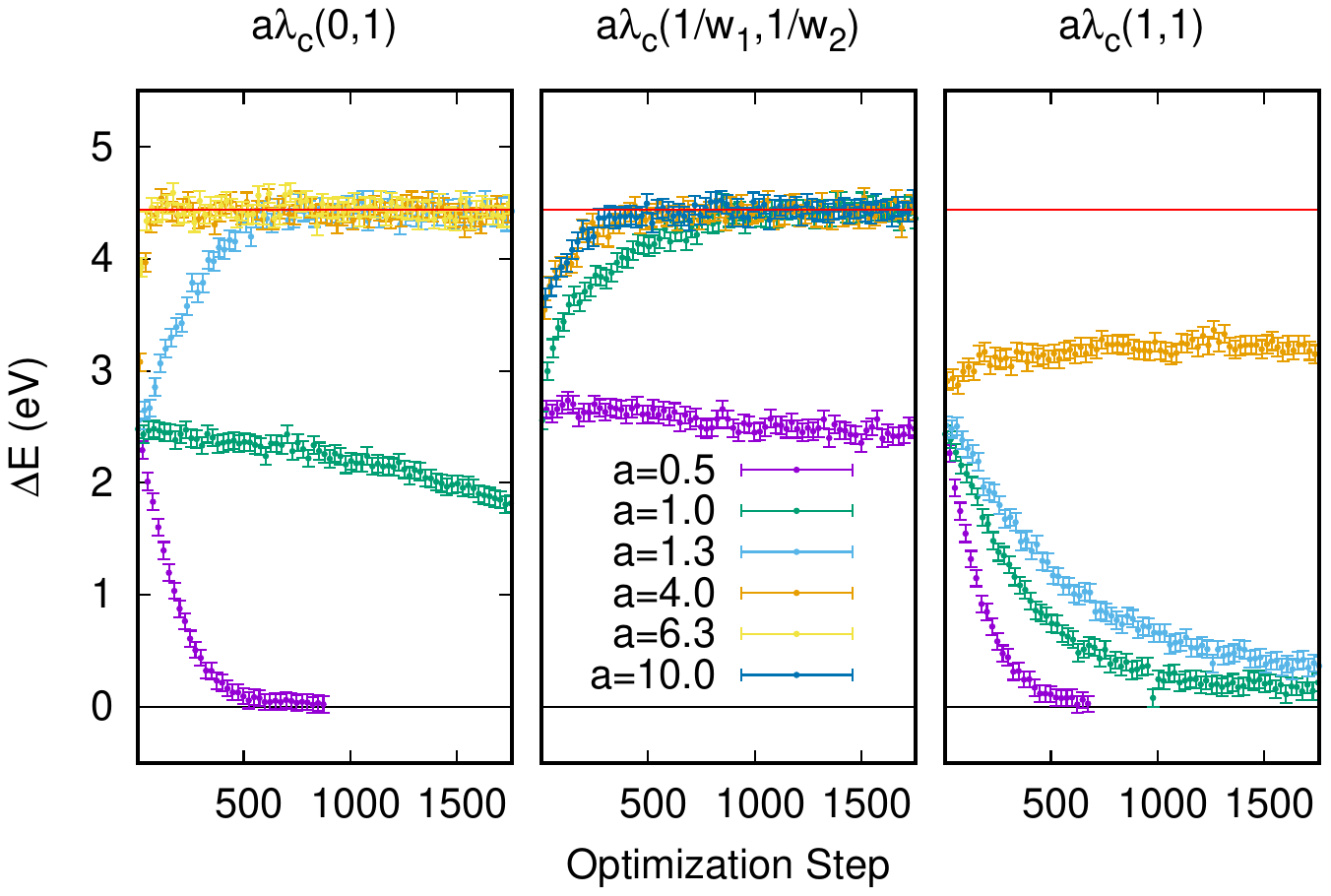}
\caption{\label{fig:testing-all} Excitation energy (eV) of nitroxyl during optimization runs started from a pre-collapsed excited-state wave function. The functionals $O_{\mathrm{<}}$ (left), $O_{\mathrm{e}}$ (or $O_{\mathrm{\neq}}$ with non-symmetric $\lambda_{IJ}$) (center), and $O_{\mathrm{\neq}}$ with symmetric $\lambda_{IJ}$ (right) are employed corresponding to different choices of $(\lambda_{12}, \lambda_{21})$, all scaled with $a\lambda_c$.}
\end{center}
\end{figure}

For the $O_{\mathrm{<}}$ functional, the orthogonality constraint is applied only to the excited state and the optimization of the ground state is thus decoupled from the excited-state one, the only influence being a lower sampling efficiency due to the use of a guiding function~\cite{Shepard2022}.
As shown in Fig.~\ref{fig:testing-all} (left), the states separate in energy for $\lambda>\lambda_c$ and, at convergence, yield the same excitation energy. In the case of $\lambda=\lambda_c$, the excited state does not recover and slowly collapses toward the ground state, and more rapidly for half the value of $\lambda_c$. There is also an acceleration in the optimization for larger and larger $\lambda$ although the relative speed up reduces after a certain value. For example, a factor of $a=6.3$ does not provide a faster convergence of the excited state than $a= 4$ and one must also take into account that a larger $\lambda$ will in principle introduce more noise into the optimization. A factor of $a=6.3$ is roughly equivalent to 1.0~Ha, which was used in Section~\ref{sec:results:validate} to validate the calculations in Shepard \textit{et al.} This verifies that $\lambda = 1.0$~Ha is more than large enough to ensure optimization to the target states in the tests we performed above.

When using the $O_{\mathrm{e}}$ functional (or equivalently $O_{\mathrm{\neq}}$ with non-symmetric $\lambda_{IJ}$), we must set two prefactors given by $\lambda_{12} = \lambda/w_1$ and $\lambda_{21} = \lambda/w_2$ (Eqs.~\ref{eq:grad:wheeler} and \ref{eq:grad:shepard})  where, again, $\lambda = a\lambda_c$ and the weights $w_1$ and $w_2$ are set as in the Appendix. Looking at Fig.~\ref{fig:testing-all} (middle), we find that the excitation energy follows the expected behavior with increasing scaling of the penalty. For $a \geq 1.0$, the states separate in energy while, for $a=0.5$, the excitation energy remains severely underestimated. The factors of $a=4.0$ and $a=10$ show a similar convergence of the excitation energy, yielding a faster separation between the states than $a=1$. 

Finally, in Figure~\ref{fig:testing-all} (right), we observe that, when the prefactors are symmetric ($\lambda_{12} = \lambda_{21}$), values of $a < 4$ do not recover the target states but lead to a collapse of the excited state. For $a=4.0$, the overlap suggests that the states are orthogonal (not shown), yet the excitation energy is far lower than expected for the optimal wave functions. 

\section{\label{sec:conclusions}Conclusions}

In our previous work~\cite{Shepard2022}, QMC calculations were performed for the double excitation energies of prototypical challenging molecules, where the wave functions were optimized in a state-specific fashion penalizing the states for having non-zero overlap to all other states. The resulting functional, named here $O_{\mathrm{\neq}}$, was chosen to have symmetric prefactors, $\lambda_{IJ}$, for the orthogonality constraints between two states so that the ordering of the states does not need to be known \textit{a priori}. The use of symmetric prefactors, however, guarantees that the states optimize to the target ones only in the limit of $\lambda_{IJ}$ going to infinity~\cite{Wheeler2024}. 

Here, we revisit our calculations imposing orthogonality only to the lower states using what we call the $O_\mathrm{<}$ functional, and optimizing all states simultaneously. We confirm the accuracy of our previous results and the predictive capability of the approach outlined therein~\cite{Dash2019,Dash2021,Cuzzocrea2020,Cuzzocrea2022} for computing excitation energies in QMC. For nitroxyl, however, the excitation energy obtained with the $O_{\mathrm{\neq}}$  functional~\cite{Shepard2022}  is somewhat lower. We attribute this discrepancy to the particularly long optimization performed with $O_{\mathrm{\neq}}$ for this molecule, which leads to a growing overlap for the chosen $\lambda$ prefactor in the orthogonality constraint. While increasing $\lambda$ delays this growth, appropriate functionals~\cite{Pathak2021,Wheeler2024} such as $O_\mathrm{<}$ or the ensemble $O_e$ variant should be employed to prevent state mixing.

Finally, we test the use of the functionals $O_{\mathrm{\neq}}$, $O_\mathrm{<}$, and $O_e$ in a state-specific optimization for nitroxyl, starting from an excited-state wave function that was intentionally collapsed toward the ground state. While $O_{\mathrm{\neq}}$ fails to recover the target states, both $O_\mathrm{<}$ and $O_e$ quickly respond to the presence of the orthogonality constraint for sufficiently large prefactors.  The availability of reliable functionals~\cite{Pathak2021,Wheeler2024}, along with effective optimization approaches, enhances the robustness of QMC for excited states, especially when combined with the use of smart multi-determinant wave functions~\cite{Dash2019,Dash2021,Cuzzocrea2020,Cuzzocrea2022,Shepard2022}.

\begin{acknowledgements}
We thank Lucas Wagner for useful discussions.
This work was in part supported by the European Centre of
Excellence in Exascale Computing TREX --- Targeting Real Chemical
Accuracy at the Exascale. This project has received funding from the
European Union's Horizon 2020 --- Research and Innovation program ---
under grant agreement no.~952165.
This work used the Dutch national e-infrastructure with the support of the SURF Cooperative using grant 
no.~NWO-2022.032.
\end{acknowledgements}

\section*{Data Availability Statement}
The data that support the findings of this study are available from the corresponding authors upon reasonable request.

\appendix*
\section{Weights for a two-state system}

As shown in Ref.~\cite{Wheeler2024}, when using the ensemble functional, $O_e$, the pairwise $\lambda_{IJ}$ should be larger than a critical value,
\begin{equation}
\lambda_{IJ} > \lambda_c^{IJ}=\left(E_I - E_J \right) \frac{w_Iw_J}{w_J-w_I},
\label{condition_lambda}
\end{equation}
where $w_I < w_J$ if $E_I> E_J$. In practice, given a set of weights $\{w_L\}$, a single critical $\lambda_c$ can be used corresponding to the maximum of the critical values. A possible choice of weights is the one that gives the same critical lambda for all pairs of states, yielding one free parameter, $\lambda_c$, and an iterative approach to determine the weights.

For a 2-state system, if we set $\lambda_c^{21} = \lambda_c = \left(E_2 - E_1 \right)$, we can rewrite Eq.~\ref{condition_lambda} as
\begin{equation}
1 = \frac{w_1w_2}{w_1-w_2},
\end{equation}
which, when used with the condition on the sum of the weights, $w_1+w_2=1$, leads to the quadratic equation,
\begin{equation}
w_1^2 + w_1 -1 = 0,
\end{equation}
which has as solutions $w_1 = (\sqrt{5}-1)/2$ and $w_2 = (3-\sqrt{5})/2$.

This can be readily generalized to the case of a 3-state system where the energies of the two excited states are equal, etc. For a 3-state system of this type, using that $w_2=w_3$ and that $w_1+w_2+w_3=1$, one obtains $w_1=(\sqrt{2}-1)$ and $w_2=w_3=(1-1/\sqrt{2})$.

\bibliography{bibliography}

\end{document}